# A Study on the Optimal Implementation of Statistical Multiplexing in DVB Distribution Systems

**Alexandru Florin Antone**
Telecommunications Department
Technical University of Cluj-Napoca
Cluj-Napoca, Romania
antonealexandru@yahoo.com

**Radu Arsinte**
Telecommunications Department
Technical University of Cluj-Napoca
Cluj-Napoca, Romania
Radu.Arsinte@com.utcluj.ro

*Abstract*—The paper presents an overview of the main methods used to improve the efficiency of DVB systems, based on multiplexing, through a study on the impact of the multiplexing methods used in DVB, having as a final goal a better usage of the data capacity and the possibility to insert new services into the original DVB Transport Stream. This study revealed that not all DVB providers are using statistical multiplexing. Based on this study, we were able to propose a method to improve the original DVB stream, originated from DVB-S or DVB-T providers. This method is proposing the detection of null packets, removal and reinserting a new service, with a VBR content. The method developed in this research can be implemented even in optimized statistical multiplexing systems, due to a residual use of null packets for data rate adjustment. There is no need to have access in the original stream multiplexer, since the method allows the implementation "on the fly", near to the end user. The proposed method is proposed to be applied in DVB-S to DVB-C translation, using the computing power of a PC or in a FPGA implementation.

## I. INTRODUCTION

DVB is already a mature technology, verified over more than 15 years in many countries. With its variants S/T/C of the first generation, and T2/C2/S2 of the second one, DVB is used by a large number of audio/video content distributors.

Broadcasters, network operators direct satellite broadcasting and other digital video services providers are seeking to increase the number of services transmitted in a fixed channel bandwidth, while maintaining high quality video. Today, for this purpose, it is used the statistical multiplexing system with a feedback between the multiplexer and encoders.

Two forms of multiplexing [1] are commonly used today: time-division multiplexing and statistical multiplexing (Fig.1).

A. Time-division multiplexing is providing a fixed amount of bandwidth for each incoming stream. The packets from each incoming stream are placed into one or more timeslots in the combined stream. Generally, this allocation can be adjusted to accommodate streams that require variable bandwidth, or in systems where the allocation cannot be changed rapidly or while the system is in use

B. Statistical multiplexing allocates bandwidth to input channels in response to their needs; high-speed channels receives a larger amount of the overall network capacity. The systems can be configured with a maximum and a minimum bit rate for each tributary stream. In the same time, a rate limit is imposed on the combined stream.

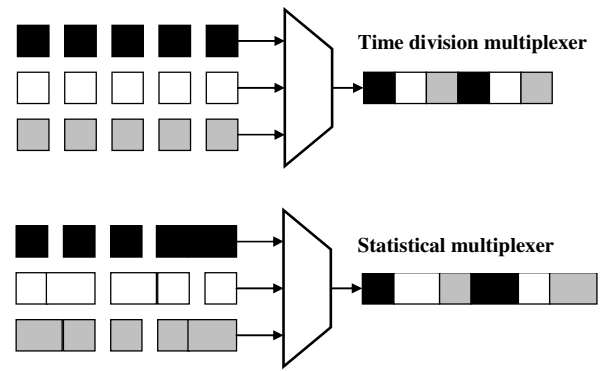

Figure 1. Multiplexing (fixed and statistical) [1]

Considering implementation of DVB as digital fixed and mobile television in the world, end-users and providers are looking for optimization in the transmission system. Optimization is desired from technical point of view and also from business point of view, distributing more attractive services, lower maintaining costs etc. Statistical multiplexing is a critical factor in mobile DVB, with the (still) reduced data capacity of the mobile network. Even the basic generations of DVB (T/S/C) are taking a benefit from statistical multiplexing, raising the audio-video streaming capacity. One of key factors in obtaining this desiderate is the statistical multiplexing. It implies a better use of network bandwidth through medium bandwidth allocation of every link connection, instead of absolute/dedicated bandwidth allocation.

Statistical multiplexing architectures in DVB allows media broadcasters the increase of efficiency in distributing digital services, thing that leads to an increase of profits along with customer satisfaction due to broad variety of audio video services. It has been noticed that in case of video coders operating in constant bit rate mode, complex video scenes don't have sufficient bandwidth and therefore are error prone (for e.g. art effect can be noticed) opposed to simple video scenes which don't require as much bandwidth. Therefore in case of simple video scenes bandwidth is wasted. Conducted studies analyzing several parallel video streams showed that simultaneous existence of complex video scenes in parallel video streams are very rare. Therefore an advantage is to use a variable coding bit rate (of such streams) using a lower coding rate for simpler video scenes and allowing a higher coding rate for other streams with complex animations, taking advantage of a higher bandwidth. Another advantage is to use the gained bandwidth (by using lower coding rates for simpler video scenes) for inserting additional services (as text or other non time-critical application). In this paper, we are focused on conducting a study of classical or statistical multiplexing, used by different service providers of DVB content, transmitted by telecommunications satellites or terrestrial, having as a final goal a better usage of channel capacity, and the possibility to offer new services in DVB networks.

## II. IMPORTANCE OF STATISTICAL MULTIPLEXING IN DVB SERVICES

The aim of Statistical Multiplexing is to increase the capacity of the multiplex taking advantage of the variability of bit rate along time of the different services when a sufficient number of them are combined into a Transport Stream. Statistical Multiplexing has been developed successfully in other DVB networks such as DVB-T, DVB-H or DVB-S.([2] [3] [4] [5]).

### A. Modes of Statistical Multiplexing

There are four modes to set encoding speed.

**Mode Variable Bite Rate (VBR)** (Fig. 2)

Coding rate is selected in accordance with a given image quality. It normally fluctuates around a certain mean value.

**Mode Capped Variable Bit Rate (Capped VBR)** (Fig.3)

Speed is selected in accordance with a specified image quality, but with the forced setting of the upper limit. As in the first mode, the speed will fluctuate within a certain range of the average, do not exceed the ceiling.

**Mode Available Bit Rate (ABR)** (Fig.4)

Encoder uses the bandwidth that is allocated to it on an fixed basis.

**Mode Constant Bit Rate (CBR)** (Fig.5)

As in the previous mode, the encoder uses the maximal data rate allocated permanently, and the remaining gaps are filled with null packets.

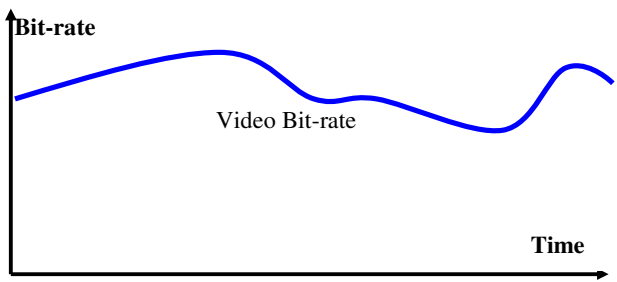

Figure 2.  Mode Variable Bit Rate (VBR)

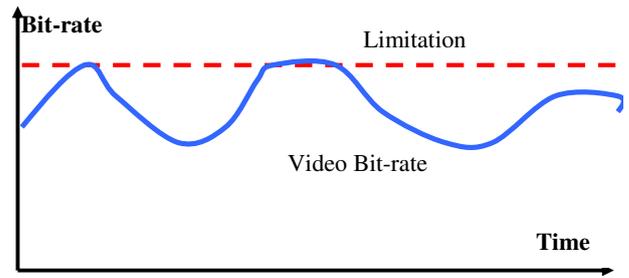

Figure 3.  Mode Capped Variable Bit Rate (Capped VBR)

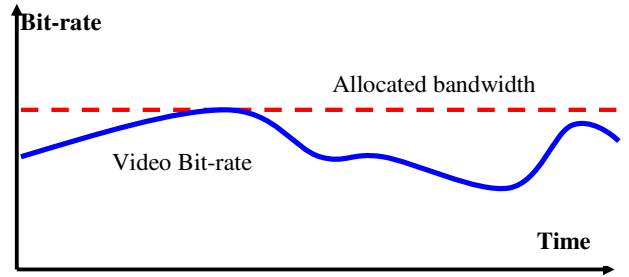

Figure 4.  Mode Available Bit Rate (ABR)

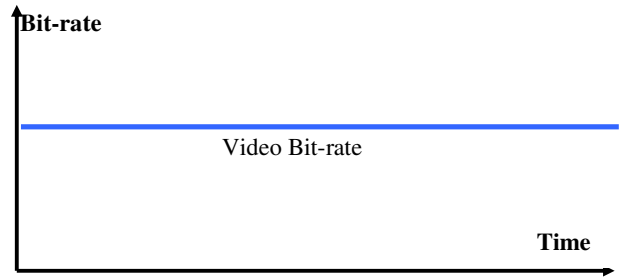

Figure 5.  Mode Constant Bit Rate (CBR)

CBR (Constant Bit Rate) is a term used when media contents has an encoding margin. This complication makes multiplexing in bursts more difficult than normal "continuous" multiplexing, since it introduces a new constraint which must be taken into account ( [1] ).

VBR (Variable Bit Rate) encoding, opposed to CBR, varies the amount of information per time segment. This means that complex parts of the encoded source are allocated with more capacity while the less complex parts are allocated with less capacity and so the perceived quality remains constant. This variation in bit rate normally fluctuates around an average bit rate which refers to the average amount of data transferred per unit of time ([1] [2] [6] [7]).

### B. Statistical multiplexing with feedback (Closed Loop)[8]

In real DVB systems the bit rate adjustment is done using the audio-video encoders in a feedback architecture. In a system with feedback, the encoders transmit the information directly about the complexity of each video sequence to the statistical multiplexing system, which assigns individual flow rates. Such a scheme is often used at the Headend, where all encoders are in the same place. The presence of feedback allows the bandwidth to be used more efficiently.

*C. Statistical multiplexing without feedback (Open Loop) [8]*

If the initial exchange of information between the encoder and multiplexer system is missing, the system is without feedback. This variant is sometimes called rate-shaping statistical multiplexing.

Reducing the flow rate in the system is achieved by the following methods:

1. Manipulation of the input compressed streams for avoiding simultaneous occurrence of high-speed peaks. Speed peaks appear at times when coders work in the load mode. The time shift input flows through their buffering to spread peaks, minimizing the peak rate of the overall flow.

2. Incoming stream is encoded at CBR, and can contain up to 10% of null packets in the multiplex, which can be removed.

3. The flow rate can be reduced by optimizing the coefficients of the quantization matrix of input compressed stream. This method is used in cases where the flow will pass on a narrower channel than the one on which it was passed in the first stage. This procedure is often performed on flows from satellite DTH packages before they are relayed to the IPTV networks. Sometimes it is called trans-rating. Systems without feedback are the typical Headend cable TV networks and IPTV, where the re-multiplexing streams are already encoded in remote broadcasting centers.

*D. Organization of the system, depending on network*

In traditional satellite networks that form MPEG-2 (or MPEG-4) packets for DVB reception, the dominating system is with feedback, where the featured encoders are connected to the multiplexing system via a reliable LAN. However, many applications are not allowing such a connection to the encoder system. Organization of the return channel can be expensive, or simply is not possible to provide the required reliability of the channel encoders and tight synchronization. At the same time, content owners increasingly are engaged in a compression of the material, creating a need for distributed statistical multiplexing.

A necessary condition for such a system is the possibility of organizing a global network of reliable channels connecting the station with all remote locations. Despite the complexity and high cost of technology, it is a realistic option for many applications, especially in those cases where recompression is necessary.

In the world of digital television, statistical multiplexing is used on a large scale as a mean of obtaining a 30-40% efficiency for a given transmission channel (terrestrial, satellite, cable); as a proprietary and accepted technology by all broadcasters.

The concept of statistical multiplexing in DVB is based on the fact that only changing or highly complex video scenes require a higher bandwidth and that the rest of the video data can be sent at a lower bit rate. The bandwidth setting on CBR encoders has to be established that these complex scenes can be transmitted with good quality, meaning that the encoders would generate on average more bandwidth than they would really require.

*E. Advantages of using Statistical Multiplexing [1]*

The achieved gain by using statistical multiplexing in digital video television may be used to enhance the video quality of the available services while keeping their number constant or transmit more video services of the same quality on the same available network transmission bandwidth.

Statistical multiplexing provides significant effectiveness in terms of operation and use of transportation resources, as well as a significant reduction in costs. It provides the following key benefits:

• Any service can be included in any package that provides flexibility and scalability;

• Simplification of content aggregation by eliminating distance limitations;

• One-step statistical multiplexing leads to a more efficient use of transport resources, optimizes content aggregation scheme and make possible to balance the load of primary and secondary (backup) system;

*F. Empirical Statistical multplexing algorihtms*

In the literature some theoretical or practical studies ([8][9]) proposed two statistical multiplexing algorithms that are based on the empirical rate-distortion prediction algorithms and equal distortion bandwidth algorithm. Based on those approaches, the bandwidth allocation performed once every GOP (group of pictures) in two stages: first, the rate-distortion functions of the next GOP is predicted for all input channels, second, an optimal bandwidth allocation is computed for every channel based on predicted rate-distortion functions and the total bandwidth so that the expected distortion of every input channel is equalized and the total bit rate is under the budget.

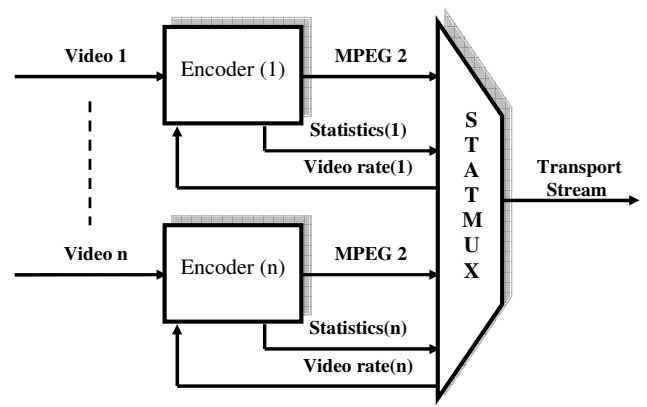

Figure 6. An example of statistical multiplexing system [2]

A statistical multiplexing algorithm can be divided and distributed on the encoders and then the multiplexor. Based on the system structure, every encoder calculates the relevant statistics of the current GOP and sends them to the multiplexer at every GOP boundary. After receiving all the statistic, the multiplexer first predicts the rate-distortion functions of the next GOPs from the corresponding statistics. Then the multiplexer assigns video rates for all the input channels. Finally, the multiplexer sends back bandwidth allocation to every encoder. Upon receiving its video rate sent back from the multiplexer, the encoders continue to compress the next GOP of video sequence using the specified video rate (as described in Fig. 5).

*G. Analytical statistical multplexing algorihtms*

Other studies ([8]) are proposing two statistical multiplexing algorithms based on the theoretical rate-distortion model. Based on the rate-distortion model ever input channel computes and sends the relevant statistic of its current frame to the multiplexer at the end of the transformation stage. After receiving the relevant statistics of the current frame for every encoder, the multiplexer first estimates the rate-distortion functions from the corresponding statistics, the, it computes the quantization scale factor for every encoder based on the estimated rate-distortion functions and the available bit budget of the current frames, so that the expected distortion of all input channels are equalized and the expected total bits (used to encode the current frames) are under the budget. Then, the multiplexer sends back calculated values to the corresponding encoders.

Upon receiving, the encoders continue the second stage of video compressing process (encoding stage), which quantizes the transformed frame with the specified values and encodes the quantized DCT (Discrete Cosine Transform) blocks and motion vectors into an MPEG-2 video stream (as illustrated in Fig. 6). The same architecture is applied even for newer standards (MPEG4, H264).

Video sequences may be considered: simple, moderately complex, complex, extremely complex. Several simulation trials ([10][11]) were conducted to evaluate overall performance of the statistical multiplexing algorithm. The multiplexer evaluates the output bit rate of each encoder (MPEG Video) and adjust (using Mquant output) the encoding algorithm to keep the Transport Stream bit rate within allowed limits.

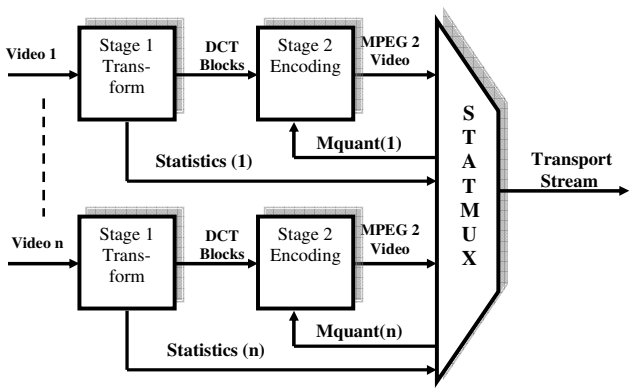

Figure 7. Simulation of a statistical multiplexing system based on modified encoders [8]

The algorithm is designed to bring the following advantages: reducing the average distortion (improving average picture quality) of a set of channels given with a fix bandwidth, an increase in the number of channels without increasing the average distortion given a fixed bandwidth, decreasing the total bandwidth required to compress the same set of video sequences without increasing its average distortion. Different bibliographic sources ([5],[6],[7]) are presenting various aspects of statistical multiplexing in DVB.

## III. EXPERIMENTAL SYSTEM

In this phase of the research, we conducted a study on the implementation of multiplexing methods, classical or statistical, used by different service providers of DVB content transmitted by telecommunications satellites or terrestrial. The research was facilitated by a set of software tools, recommended in different studies ([12] [13] [14]). Our previous experience in this field was particularly useful.

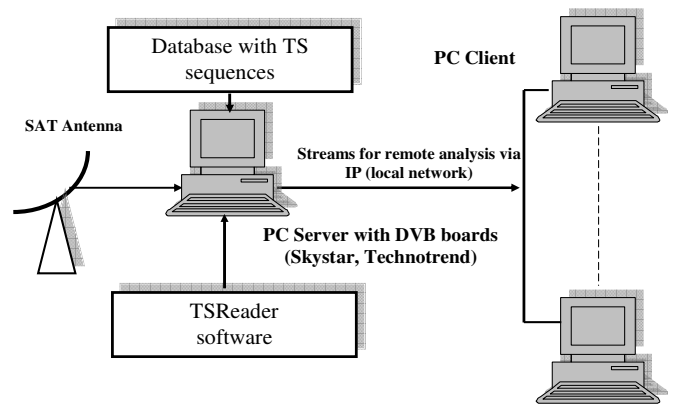

Figure 8. The system for DVB stream analysis [12]

The goal of the experiments was to establish the frequency of use of static or statistical multiplexing in different DVB applications and to establish the provision of data capacity offered by null packets, to decide the best architecture of the re-multiplexer.

The system (described in detail in [12], briefly in Fig. 8) is composed from a PC equipped with one or more DVB boards or modules. This allows both real-time analysis of DVB streams, or file based analysis.

The system has the possibility to stream the desired content in the local network, making possible to implement simultaneous analysis of different parameters. The system has also an educational value, being used in various experiments in university laboratory and research projects.

*A. The software - TSReader*

The TSReader [14] is a transport stream analyzer (TS) decoder, recorder and stream manipulator for MPEG-2 (not exclusively) systems. The application provides to the user information on a multiplex and can be very useful for identifying errors and inconsistencies of TS sources. The general view of the program with the different windows is presented in Fig. 8.

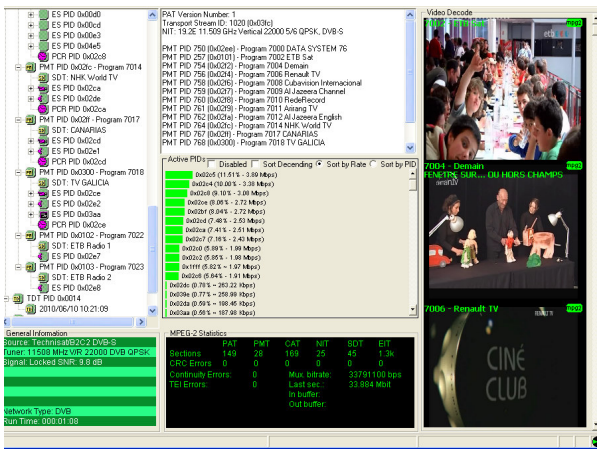

Figure 9. A general view of TSReader GUI [17]

*B. DVB Stream Analysis*

We have as a model the study described in reference [10] and implemented using professional equipment. Our goal was to obtain useable results with less expensive equipment. In this section we will present the results with constant references to [10], study accessible on Internet.

The first analysis was performed using a fragment of the TS captured from transponder 89 (12168.00MHz - H) from Astra 1 satellite. The content of the multiplex analyzed is presented in Fig. 10.

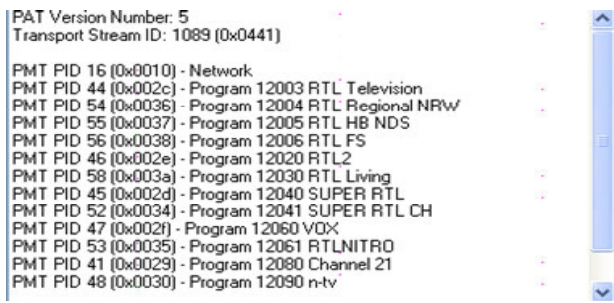

Figure 10. The structure of RTL multiplex delivered by TSReader

The analysis presents the contribution of each partial stream to the entire multiplex, with different colors like in Fig.11. At the first glance the analysis of the multiplex usage, presented in Fig. 11, is showing that the multiplexing method employed seems to be non-statistical. This is an effect of the fact that an averaging option is applied in TSReader settings.

To prove that, next diagram (Fig. 12) is presenting a diagram realized with both options (real-time - left and averaged- right). First half of diagram is plotted with real-time option, the last part is employing averaged information. The advantage of averaging function is that it is possible to evaluate precisely the average value of the data rate for each component.

In this case of statistical multiplexing the weight of packets containing null information (upper layer – white color) is still high. Basically, more than 3% of the bandwidth is not used.

Statistical multiplexing allows eventually to add additional channels in the same multiplex. The non-averaged option allows to see the peaks of the information rate and adopt the appropriate measures to handle that.

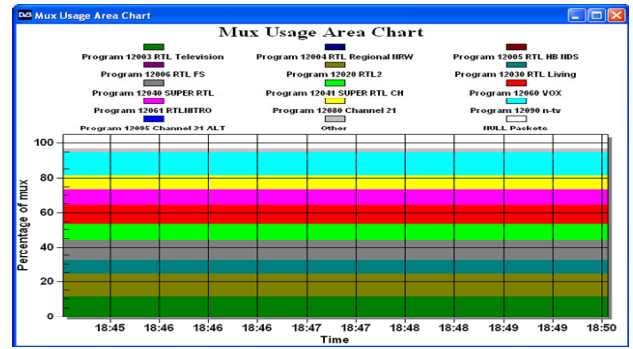

Figure 11. Mux usage stacked area chart for RTL mux (Averaging)

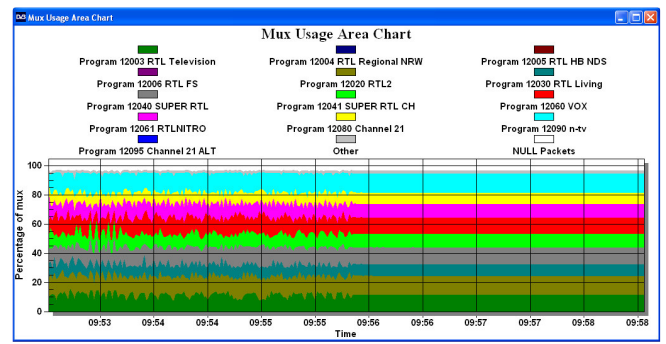

Figure 12. Real-time and averaged diagram of mux usage area chart (RTL mux)

PID usage from Fig. 13 presents the actual values for each stream. PID usage display option of TSReader can be used to establish exactly the values to be allocated for a possible additional service insertion in the DVB transport stream, as we will propose in this paper.

Most providers are using, despite technical complications related to the implementation, statistical multiplexing but some are continuing to use non-statistical multiplexing. A detailed analysis for most transponders of Astra fleet (19,2°E) found a general use of statistical multiplexing.

The second example (Fig. 14) (transponder 65 - 11719.50 H) shows an encrypted multiplex with statistical multiplexing. In this example programs like Sky Bundesliga or Sky Sports 1 are programs with a large variability of the data rate, where statistical multiplexing brings a real benefit. The yellow stream (Program 2691 NDS Application) contains no real video information, but the application organizing the entire stream. A brief research on NDS revealed that this is a middleware application of Sky. In this case statistical multiplexing cannot be applied, since the redundancy of data is probably zero. Statistical multiplexing is ideal for services with a large variability in video data rate, and sports channels are the ideal candidates.

In Table I we are gathering the results of the measurements for different video streams of the multiplex (within the observation window - approx. 2 minutes). It is obvious that only VBR allocation allows to have a large number of channels, the maximum bit-rate on a CBR basis would require a larger space than allowed by the transponder. Note that the table is not including the audio bitrates and applications bitrates. In this case the difference would be more significant. So, theoretically the utilization of the channel varies from a "surplus" of capacity of 24Mbps to a "lack" of capacity of 9 Mbps. Of course the two extreme situations are not met simultaneously, the variation of audio/video streams smoothening the average bit-rate.

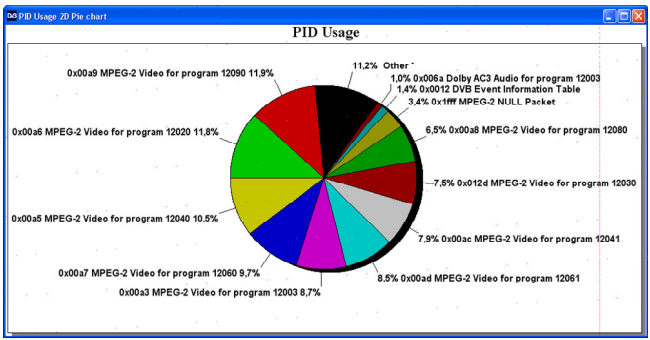

Figure 13. Pie chart mux usage of an RTL multiplex

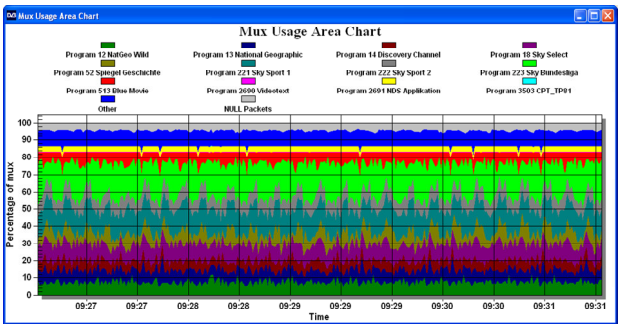

Figure 14. Mux usage of an encrypted multiplex

To illustrate that non-statistical multiplexing is still used, we are presenting in Fig. 15 an example, using a DTV multiplex from source [15]. In this case we have a mix of HDTV and SDTV services, the HDTV service occupies 70% of the total capacity of 19Mbps. Program 3 is active, but represents a static image (logo), so the data rate is extremely low. For the presented multiplex, with both settings of TS Reader (real-time and averaged) the results are the same. Non-statistical multiplexing used in this example is offering a large provision for additional service insertion, about 10% of the full capacity, as seen clearly in multiplex usage diagram.

The conclusion is that systems with large data capacity seems less attractive for broadcasters to implement statistical multiplexing. But such situations are rare, since the final goal for all program providers is to maximize data throughput. So, cases like those described in Fig.15 will be probably less frequent in the future.

Despite of that, DVB-T version seems to be suitable for statistical multiplexing. The source [16], contains an extended analysis of different DVB-T transponders, from all parts of the world, and many stations presented there are employing statistical multiplexing. An example, obtained with the same TSReader program is presented in Fig. 16. An explanation is that the capacity of DVB-T multiplex is significantly reduced (normally around 20Mbps) compared with DVB-S (starting at 38Mbps), and any method improving de data rate is welcomed.

TABLE I. BITRATES OF VIDEO STREAMS FOR SKY MULTIPLEX

| Pro-gram | Video bit-rate (Mbps) | | |
|---|---|---|---|
| | *Channel name* | *Max* | *Min* |
| 1. | NatGeoWild | 3.5 | 1 |
| 2. | Nat. Geographic | 3.5 | 0 |
| 3. | Discovery | 3.5 | 1.2 |
| 4. | Sky Select | 5.2 | 1.8 |
| 5. | Spiegel Geschichte | 5.8 | 0.6 |
| 6. | Sky Sports 1 | 7.8 | 3.8 |
| 7. | Sky Sports 2 | 4.6 | 1.4 |
| 8. | Sky Bundesliga | 10 | 3 |
| 9. | Blue Movie | 5 | 0.6 |
| . | **Total** | 48.9 | 13.4 |
| | **Total available on transponder** | 38 | 38 |
| | Difference | -9.1 | 24.6 |

This program, in different versions, is a simple to professional MPEG-2 stream analyzer that analyzes digital TV feeds from satellite, cable, terrestrial or IP sources and displays information about the programs and tables carried in a multiplex. We are presenting the results of the evaluation of multiplexing methods employed in different broadcasting networks, using TSReader. This provides more information about the multiplex for a specific transponder: Number, name, PID (Packet Identifier-ID packet) multiplex TV programs (MPEG-2 elementary streams TS), tables MPEG-2 TS flow (PAT, PMT, CAT, NIT, SDT, EIT), during the reception of multiplex standard parameters presented above (SNR, frequency of the transponder, type of polarization, modulation, symbol rate, useful bit rate). In addition to these parameters we can estimate the subjective quality of TV programs (the multiplex) via its Video Decode section.

Even in the case of statistical multiplexing, we still have a fair amount of data (concentrated in null –packets) unused in the multiplex. In our experiments this value is found to be around 3%. Table II presents the results of a study regarding the null packets percentage on several statistically multiplexed streams, captured on the same 19 degrees Astra Satellites package. The explanation of null-packets still existent in the multiplex, is that these are needed to ensure the constant bit rate value of the multiplex. The results are comparable with the results of table 2 from [10].

Our idea, still in an early stage of implementation, is to reuse this capacity for applications not requiring a fixed bit rate, or a real-time behavior. Rewriting the PMT (program map table) with the new service and replacing the null-packets with the content of the new service might add useful features to the entire multiplex. A practical conclusion is that TSReader must be correctly configured to provide useful results in the investigation of multiplex services.

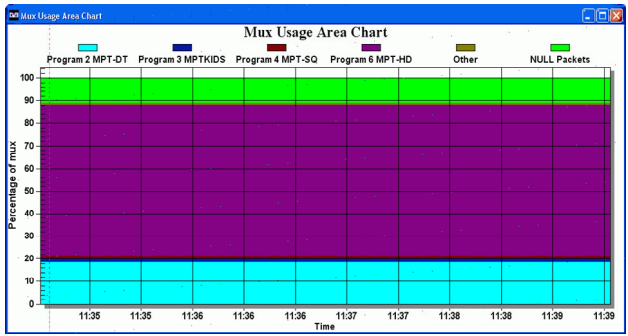

Figure 15. An example of non-statistical multiplexing

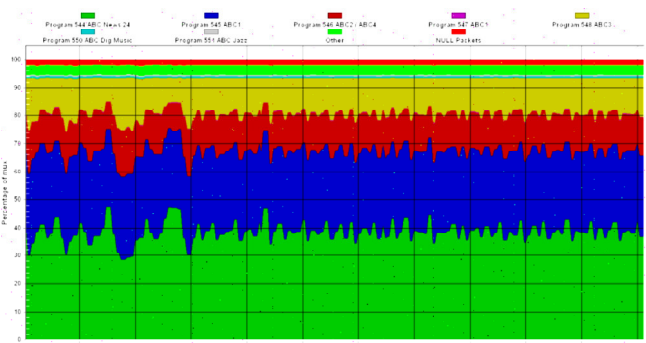

Figure 16. Multiplex usage of a DVB-T Australian broadcaster [16]

The proposed re-multiplexing architecture is designed to modify the original content of the multiplex, without needing access to the initial multiplexer of the original broadcaster. We are proposing an updated architecture (described briefly in Fig. 17), with a null packet detector, a content inserter, and finally a classical multiplexer, as seen in the mentioned diagram.

TABLE II. PERCENTAGE OF NULL PACKETS IN MULTIPLEX

| Mux No. | Bitrates (Mbps) | | | |
|---|---|---|---|---|
| | *Transponder* | *Total* | *Null* | *%* |
| 1. | Sat1/Kabel1 | 38 | 0.5 | 2 |
| 2. | Sky | 38 | 1.8 | 4.7 |
| 3. | RTL | 38 | 1.5 | 3.9 |

The process implies also a new structure for PAT (Program Allocation Table regenerated by the appropriate block), and eventually the PMT (Program Map Table) if the new service has an audio/video content. It is also possible to adopt an proprietary format for inserted data. In this case an appropriate client should be implemented.

The system is re-multiplexing the incoming stream adding new services, without overloading the output channel.

In this phase of the research the implementation is purely software, with a clear limitation of the stream bandwidth. The demonstration program performs only file to file re-multiplexing.

The following implementations will try to use the enhanced processing capacity of FPGA's or DSP, in an attempt to achieve a real-time behavior.

This proposed structure (still in development) need an extended set of experiments to establish the best parameters of each block.

In this study, we have presented an analysis of multiplexing methods in DVB services, having as a final goal to develop methods for a better usage of the channel capacity, opening possibilities to offer new services in DVB networks. This initial study proved to be an important step in our research, since the most technical problems of multiplexing, and the use of the real capacity in DVB networks has not received enough attention in previous literature.

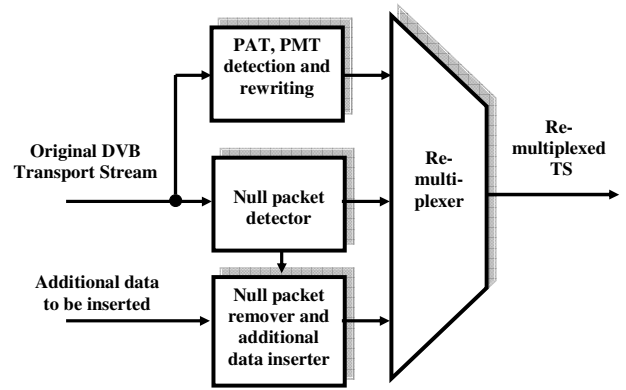

Figure 17. Block diagram of the DVB service inserter proposed

IV. CONCLUSIONS

We have presented in this work a practical (experimental) study to establish the parameters employed by statistical multiplexers in DVB systems. The purpose of the research is to establish if the transport stream delivered can be further be processed on clients side to add new services or enhance the existing services.

Our early studies (described in [13] and [17]) proved that a DVB multiplex, normally optimized for broadcasting, still has possibilities to be used in a more efficient way. There are at least two methods to improve that.

First method, emphasized in this enhanced study, is the use of statistical multiplexing. Statistical multiplexing is widely used in the world of digital television as a means to obtain an efficiency boost of 30-40% for a given transmission channel (terrestrial, cable or satellite) and the technology it is mastered and accepted by broadcast operators. The principle is based on the observation that only changes or highly complex scenes require a higher band width and the rest of the image data can be sent with a lower bit rate. Setting the bandwidth of a CBR encoder must satisfy the requirement for complex scenes to be sent with high quality, i.e. encoders to generate more bandwidth than actually required. Statistical multiplexing requires replacement of old equipment with new one implementing the new multiplexing algorithm, which could be costly for small DVB operators.

The second method, proposed by our present paper and reference [17], is to reuse the capacity offered by null packets, and reinsert an additional service, less sensitive to QoS parameters, as for example, static web pages (informational pages, public interactive services, newscast services). This has the advantage that the equipment and software needed is less complex, avoiding the implementation of adjustable MPEG encoders with VBR. Our first experiments proved that a normal PC is able to perform this task. This method is intended to be applied first in DVB-S to DVB-C translation performed in specialized equipments (Headend). The next phases in our research project will continue to investigate the actual implementation of the proposed methods, to find the best solution. This method seems to be an ideal DVB content improvement method for small Cable TV operators or Internet providers. Similar studies are performed by companies, for commercial purposes in transmission systems optimization, like in the reference [18].

A different direction of the research will be to investigate the impact of statistical multiplexing of DVB services through distribution in IP based networks. According to different studies ([9]) the use of statistical multiplexing in IP-based networks, distributing DVB, has many advantages. Statistical multiplexing IP provides significant effectiveness in terms of operation and use of transportation resources, as well as a significant reduction in costs. It provides (as claimed in [9]) the following key benefits:

- Any service can be included in any package that provides flexibility and scalability;
- Simplification of content aggregation by eliminating distance limitations;
- One-step statistical multiplexing more efficient use of transport resources, optimizes content aggregation scheme and make it possible to balance the load primary and secondary (backup) system;
- Using the Ethernet-equipment simplifies the system and reduces its cost;
- An increase in flexibility and reliability of backup services;
- self-managed infrastructure simplifying the system management.

Our future work, intended to be performed in the next phase, involving the streaming of DVB information in local networks, will try to prove the claimed advantages in different DVB to IP conversion configurations.

ACKNOWLEDGMENT

This paper was supported by the project "Improvement of the doctoral studies quality in engineering science for development of the knowledge based society-QDOC" contract no. POSDRU/107/1.5/S/78534, project co-funded by the Sectorial Operational Program Human Resources 2007-2013.

REFERENCES

[1] Wes Simpson - Video Over IP: IPTV, Internet Video, H.264, P2P, Web TV, and Streaming: A Complete Guide to Understanding the Technology, Focal Press, 2008
[2] A Lopez, J. Mas, G. Fernandez, „Comparing static and statistical multiplexing in DVB-H", Proceedings of IEEE International Symposium on Broadband Multimedia Systems and Broadcasting BMSB '09, 2009, pp.1-4
[3] C. Akamine; Y. Iano, G. de Melo Valeira, G. Bedicks, "Re-Multiplexing ISDB-T BTS Into DVB TS for SFN,", IEEE Transactions on Broadcasting, vol.55, no.4, pp.802,809, Dec. 2009
[4] M. Rezaei, I. Bouazizi, M. Gabbouj, "Joint Video Coding and Statistical Multiplexing for Broadcasting Over DVB-H Channels", IEEE Transactions on Multimedia, Vol.10, No.8, December 2008, pp.1455-1464
[5] Mehdi Rezaei,. "Video streaming over DVB-H.", Mobile Multimedia Broadcasting Standards, Springer US, 2009. pp.109-131.
[6] Boris Felts, Jean Kypreos, Thomas Guionnet, "Envivio IP-based Statistical Rate Control", Envivio, White Paper, Dec.2006 http://www.envivio.com/files/white-papers/IPStatRate_Final.pdf [last accessed on 5th February 2013]
[7] Aldo Cugnini, "Transition to Digital Handbook-Multiplexing", Broadcast Engineering World, December 2010, pp.16-18
[8] Qiang Wu, „Statistical Multiplexing for MPEG-2 Video Streams", Master Thesis, MIT 1999, pp. 33-43, pp. 69-82
[9] * * *, White paper, „Open Statistical Multiplexing Architecture for Mobile TV", White paper, UDCasts, March 2007 http://www.udcast.com/ [last accessed on 3th November 2012]
[10] Dr. Manfred Kuhn, Dr. Joschen Antkowiak, "Statistical multiplexing – What does it mean for DVB-T", FKT Fachzeitschrift für Fernsehen, Film und elektronische Medien – April 2000
[11] Hamed Ahmadi Aliabad, Sandro Moiron, Martin Fleury and Mohammed Ghanbari, „No-reference H.264/AVC Statistical multiplexing for DVB-RCS" , Lecture Notes of the Institute for Computer Sciences, Social Informatics and Telecommunications Engineering, Volume 43, 2010, pp 163-178
[12] Radu Arsinte, Eugen Lupu, Simina Emerich, "A Generic Platform to Study the Basic Aspects of DVB-IPTV Conversion Process", ISSCS 2011- Proceedings of International Symposium on Signals, Circuits and Systems, Iasi, Romania , June 30 - July 1, 2011, pp.65-68
[13] Alexandru Antone, Radu Arsinte, „An experimental study of quality analysis methods in DVB-S/DVB-S2 systems", Acta Technica Napocensis- Electronics & Telecommunications, no.4/2010, pp.7-13
[14] http://www.tsreader.com/legacy/ [Last accessed on 3th February 2013]
[15] http://www.dododge.net/roku/ts-samples.html [Last accessed on 1th November 2012]
[16] http://igorfuna.com/ [last accessed on 1th November 2012]
[17] A.F. Antone, R. Arsinte - "Investigating the Impact of Statistical Multiplexing in DVB Systems - A Practical Study for DVB Services Improvement", The 1st Virtual International Conference on Advanced Research in Scientific Areas (ARSA-2012) Slovakia, December 3 - 7, 2012, pp.2005-2010
[18] Ken McCann, Adriana Mattei, "Local Television Capacity Assessment", Zetacast. Report, 2012